\documentclass[10pt,twocolumn,letterpaper]{article}

\usepackage{iccv}
\usepackage{times}
\usepackage{epsfig}
\usepackage{graphicx}
\usepackage{amsmath}
\usepackage{amssymb}
\usepackage{booktabs}
\usepackage{float}
\usepackage{bibunits}

\makeatletter
\@namedef{ver@everyshi.sty}{}
\makeatother
\usepackage{tikz}
\usepackage{soul}

\usepackage[mode=buildnew]{standalone}
\usepackage[breaklinks=true,bookmarks=false]{hyperref}

\usepackage{tikz}
\usepackage{pgfplots}
\usepackage{sansmath}
\usepackage{color}

\usetikzlibrary{positioning}
\usetikzlibrary{spy}
\usetikzlibrary{calc}

\pgfplotsset{compat=1.16}
\pgfplotsset{
  tick label style={font=\footnotesize\sansmath\sffamily},
  every axis label={font=\sansmath\sffamily},
  legend style={font=\sansmath\sffamily},
  label style={font=\sansmath\sffamily},
  title style={align=center,font=\sansmath\sffamily},
  every axis plot/.style={mark=none,line width=1pt},
}

\pgfplotsset{select coords between index/.style 2 args={
    x filter/.code={
        \ifnum\coordindex<#1\fi
        \ifnum\coordindex>#2\fi
    }
}}

\definecolor{color1}{RGB}{100,143,255}
\definecolor{color2}{RGB}{120,94,240}
\definecolor{color3}{RGB}{220,38,127}
\definecolor{color4}{RGB}{254,97,0}
\definecolor{color5}{RGB}{255,176,0}
\definecolor{color6}{HTML}{42be65}  
\definecolor{color7}{HTML}{044317}  
\definecolor{color8}{HTML}{08bdba}  

\colorlet{colorours}{color6}
\colorlet{coloroursddim}{color7}
\colorlet{coloroursrf}{color8}
\colorlet{colormr}{color1}
\colorlet{colorhific}{color2}
\colorlet{colorpq}{color3}
\colorlet{colorqc}{color4}

\usepackage{bm}

\def\1{\bm{1}}

\def\veps{{\bm{\epsilon}}}

\def\vm{{\bm{m}}}

\def\vv{{\bm{v}}}

\def\vx{{\bm{x}}}

\def\vz{{\bm{z}}}

\iccvfinalcopy

\def\iccvPaperID{6135}

\ificcvfinal\fi

\defaultbibliographystyle{ieee_fullname}
\defaultbibliography{references}

\title{High-Fidelity Image Compression with Score-based Generative Models}

\renewcommand*{\thefootnote}{\fnsymbol{footnote}}

\author{%
  Emiel Hoogeboom\footnotemark[1] \\
  Google Research \\
  Amsterdam, Netherlands \\
  \texttt{emielh@google.com} \\
  \and
  Eirikur Agustsson \\
  Google Research \\
  Reykjavík, Iceland \\
  \texttt{eirikur@google.com}
  \and
  Fabian Mentzer \\
  Google Research \\
  Zürich, Switzerland \\
  \texttt{mentzer@google.com} \\
  \and
  Luca Versari \\
  Google Research \\
  Zürich, Switzerland \\
  \texttt{veluca@google.com} \\
  \and
  George Toderici \\
  Google Research \\
  Mountain View, USA \\
  \texttt{gtoderici@google.com} \\
  \and
  Lucas Theis\footnotemark[1] \\
  Google Research\\
  London, UK \\
  \texttt{theis@google.com} \\
}

\begin{document}

\begin{bibunit}

\maketitle

\footnotetext[1]{Equal contribution.}
\renewcommand*{\thefootnote}{\arabic{footnote}}

\begin{abstract}
    Despite the tremendous success of diffusion generative models in text-to-image generation, replicating this success in the domain of image compression has proven difficult. In this paper, we demonstrate that diffusion can significantly improve perceptual quality at a given bit-rate, outperforming state-of-the-art approaches PO-ELIC \cite{he2022poelic} and HiFiC \cite{mentzer2020hific} as measured by FID score. This is achieved using a simple but theoretically motivated two-stage approach combining an autoencoder targeting MSE followed by a further score-based decoder. However, as we will show, implementation details matter and the optimal design decisions can differ greatly from typical text-to-image models.
\end{abstract}

\section{Introduction}

\begin{figure*}[t]
    \centering
    \input{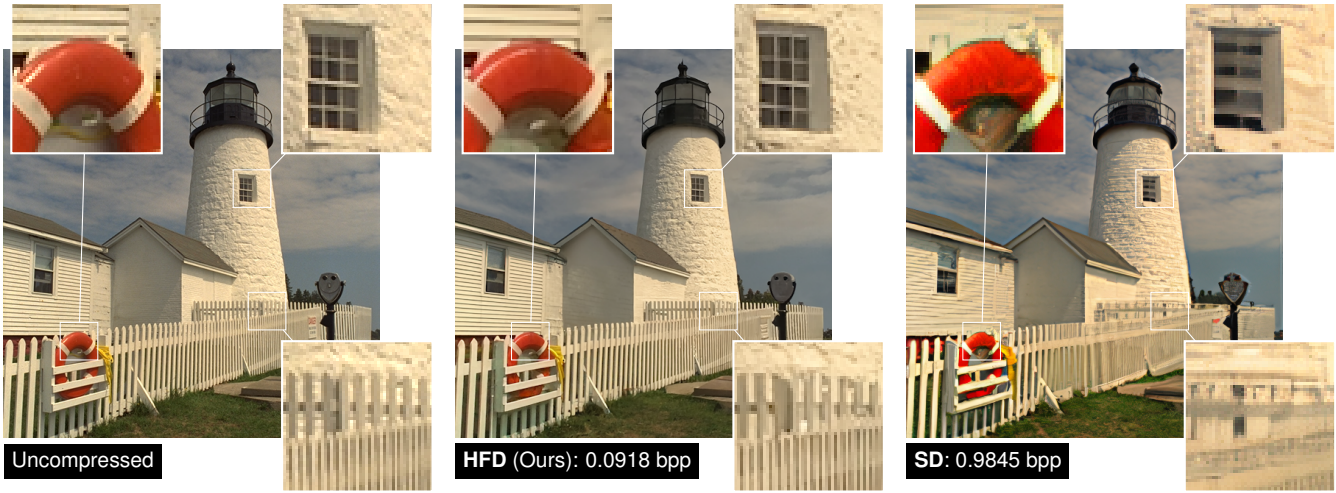}
    \caption{Popular text-to-image models struggle to reproduce fine details. As a simple baseline, we used the upsampler of Stable Diffusion \cite{rombach2022sd} applied to a $4\times$ downsampled image ($192 \times 128$ pixels). When encoded losslessly as a PNG, the downsampled image is roughly 48 kB in size (or 0.9845 bits per pixel of the full-resolution image). When encoded with JPEG (4:2:0, QF=95) \cite{itu1992jpeg}, the approach still requires 0.2635 bpp. Similar results were obtained with Imagen \cite{saharia2022imagen} and when additinally conditioning on text (Appendix~A). The example photo is from the widely used Kodak Photo CD \cite{kodak}.}
    \label{fig:text-to-image}
\end{figure*}

Diffusion \cite{sohl2015diff,ho2020ddpm} and related score-based generative models \cite{song2019generative,song2021score,song2021ddim} had an outsized impact in several domains requiring image generation, most notably in text-to-image generation \cite{aditya2022dalle2,rombach2022sd,saharia2022imagen}.
Dhariwal \etal~\cite{dhariwal2021diffusion} demonstrated that diffusion models can outperform generative adversarial networks (GANs) \cite{goodfellow2014gan} for unconditional image synthesis, and they have also been shown to outperform GANs in some image-to-image tasks such as colorization \cite{saharia2021palette} or super-resolution of faces \cite{saharia2021sr3}. It is thus surprising that score-based generative models have not yet displaced GANs for the task of image compression, performing worse or roughly on par with the GAN-based approach HiFiC~\cite{mentzer2020hific} on high-resolution images \cite{yang2022diff,ghouse2023dirac}, despite their typically higher computational cost. In line with these results, we find that trying to repurpose text-to-image models for the task of image compression does not yield good results. For instance, using Stable Diffusion \cite{rombach2022sd} (\textbf{SD}) to upscale a downsampled image either produces reconstructions which do not faithfully represent the input or which contain undesirable artefacts (Fig.~\ref{fig:text-to-image}).

\begin{figure*}[t]
    \centering
    \input{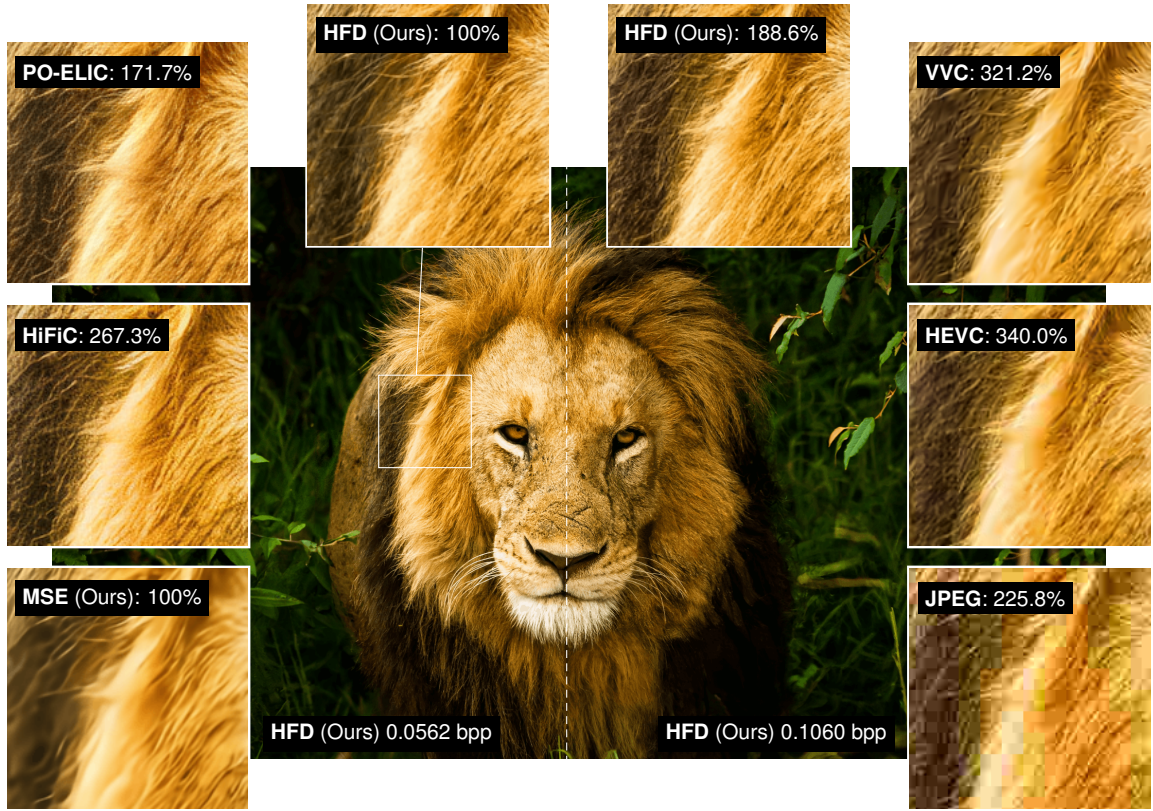}
    \caption{Illustrative example where state-of-the-art approaches based on generative adversarial networks, such as HiFiC~\cite{mentzer2020hific} or PO-ELIC~\cite{he2022poelic}, produce noisy artefacts typical for GANs (best viewed zoomed in). In contrast, our diffusion-based approach produces pleasing results down to extremely low bit-rates. Bit-rates are expressed relative to the bit-rate of our low-rate model (0.0562 bpp). MSE refers to a model similar to ELIC \cite{he2022elic} whose outputs are fed into our generative decoder. VVC and HEVC reconstructions were obtained using the reference implementation (VTM and HM, respectively). For JPEG we used 4:2:0 chroma subsampling. The photo is from the CLIC22 test set \cite{clic2022}.}
    \label{fig:hero}
\end{figure*}

In this work we tune diffusion models for the task of image compression and demonstrate that score-based generative models can achieve state-of-the-art performance in generation realism, outperforming several recent generative approaches to compression in terms of FID. For a qualitative comparison, see Fig.~\ref{fig:hero}. Our method is conceptually simple, applying a diffusion model on top of a (pre-trained) distortion-optimized autoencoder. However, we find that the details matter. In particular, FID is sensitive to the noise schedule as well as the amount of noise injected during image generation. While text-to-image models tend to benefit from increased levels of noise when training on high-resolution images \cite{hoogeboom2023simple}, we observe that reducing the overall noise of the diffusion process is beneficial in compression. Intuitively, with less noise the model focuses more on fine details. This is beneficial because the coarse details are already largely determined by the autoencoder reconstruction. In this paper, we explore two closely related approaches: 1) diffusion models which have impressive performance at the cost of a large number of sampling steps, and 2) rectified flows which perform better when fewer sampling steps are allowed.

\section{Related work}

Ho \etal \cite{ho2020ddpm} described a compression approach relying on a combination of diffusion and reverse channel coding techniques \cite{havasi2018miracle,theis2022algorithms} and considered its rate-distortion performance. Theis \etal \cite{theis2022diff} further developed this approach and demonstrated that it outperforms HiFiC \cite{mentzer2020hific} on $64 \times 64$ pixel images. However, while this approach works well, it is currently not practical as it requires efficient communication of high-dimensional Gaussian samples---an unsolved problem---so that these papers had to rely on theoretical estimates of the bit-rate.

Yang and Mandt \cite{yang2022diff} described an end-to-end trained approach where the decoder is a diffusion generative model conditioned on quantized latents. The model was evaluated on medium-sized images of up to $768 \times 768$ pixels using a large number of objective metrics and found it to perform either somewhat better or worse than HiFiC \cite{mentzer2020hific}, depending on the metric. Here, we focus on a smaller set of objective metrics since many metrics are known to be poorly correlated with perceptual quality when applied to neural methods \cite{ledig2017srgan}. Additionally, we extend our method to higher-resolution images (e.g., CLIC) and compare to more recent state-of-the-art neural compression methods. A qualitative comparison between the two approaches is provided in Fig.~\ref{fig:yang_mandt}.

An alternative approach was proposed by Ghouse \etal \cite{ghouse2023dirac}. Closely related to our approach, they first optimize an autoencoder for a rate-distortion loss, followed by training a conditional diffusion model on the autoencoder's output. The authors found that this approach peforms worse than HiFiC in terms of FID despite reporting better performance than an earlier version of Yang and Mandt's \cite{yang2022diff} model. Going further, we find that by improving the noise schedule and sampling procedure, we can achieve significantly better FID scores than HiFiC when training a diffusion model, even outperforming very recent state-of-the-art methods.

Saharia \etal \cite{saharia2021palette} explored a variety of applications for diffusion models, including artefact removal from JPEG images. However, JPEG \cite{itu1992jpeg} is known to produce relatively high bit-rates even at its lowest settings (compared to state-of-the art neural compression methods) and the authors did not compare to neural compression approaches.

Currently, state-of-the-art approaches for generative image compression are based on decoders trained for adversarial losses \cite{goodfellow2014gan}. Noteably, HiFiC has proven to be a very strong baseline \cite{mentzer2020hific}. A similar approach named PO-ELIC \cite{he2022poelic} won the most recent Challenge on Learned Image Compression (CLIC22), but using a more advanced entropy modeling approach and encoder and decoder architectures which followed ELIC \cite{he2022elic}. Similarly, we will use an autoencoder based on ELIC. 

\begin{figure}[t]
    \input{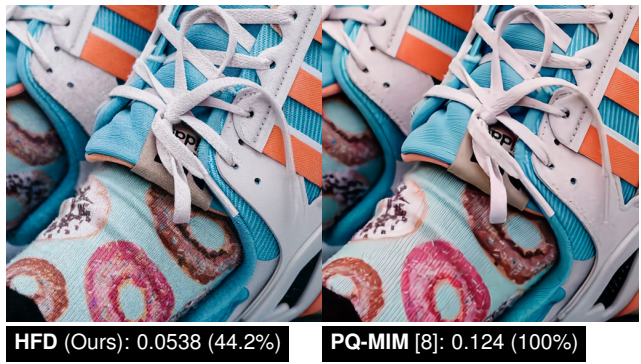}
    \caption{Qualitative comparison with the recent approach of Alaaeldin \etal \cite{alaaeldin2022vq}. We find that our approach yields significantly sharper
    reconstructions even when using a fraction of the bit-rate. Numbers indicate bits per pixel for the entire image, which is provided in Appendix~C.}
    \label{fig:pq_mim}
\end{figure}

Alaaeldin \etal \cite{alaaeldin2022vq} recently reported better FID scores than HiFiC using a combination of vector quantization, LPIPS, and adversarial training. However, as observed by the authors, reconstructions tend to be smooth and do not preserve details well (Fig.~\ref{fig:pq_mim}). This highlights the limitations of FID when comparing fine details, especially when training uses network-based loss such as LPIPS. Also recently, Agustsson \etal \cite{agustsson2022mr} described another GAN based approach which combines learnings from HiFiC \cite{mentzer2020hific} and ELIC \cite{he2022elic} to outperform HiFiC in terms of PSNR and FID on the CLIC20 and MS COCO 30k datasets, while also having controllable synthesis. 

We note that many other papers have explored variations of autoencoders trained for adversarial or perceptual losses \cite{rippel2017waveone,santurkar2018generative, agustsson2019generative, Patel_2021_WACV, huang2023towards, Wu_2020_WACV} but focus our comparisons on the recent state-of-the-art methods PO-ELIC \cite{he2022poelic} and the ``multi-realism'' approach (MR) of Agustsson \etal \cite{agustsson2022mr}.

\section{Background}
\begin{figure}[t]
\centering
    \includegraphics[width=.49\textwidth]{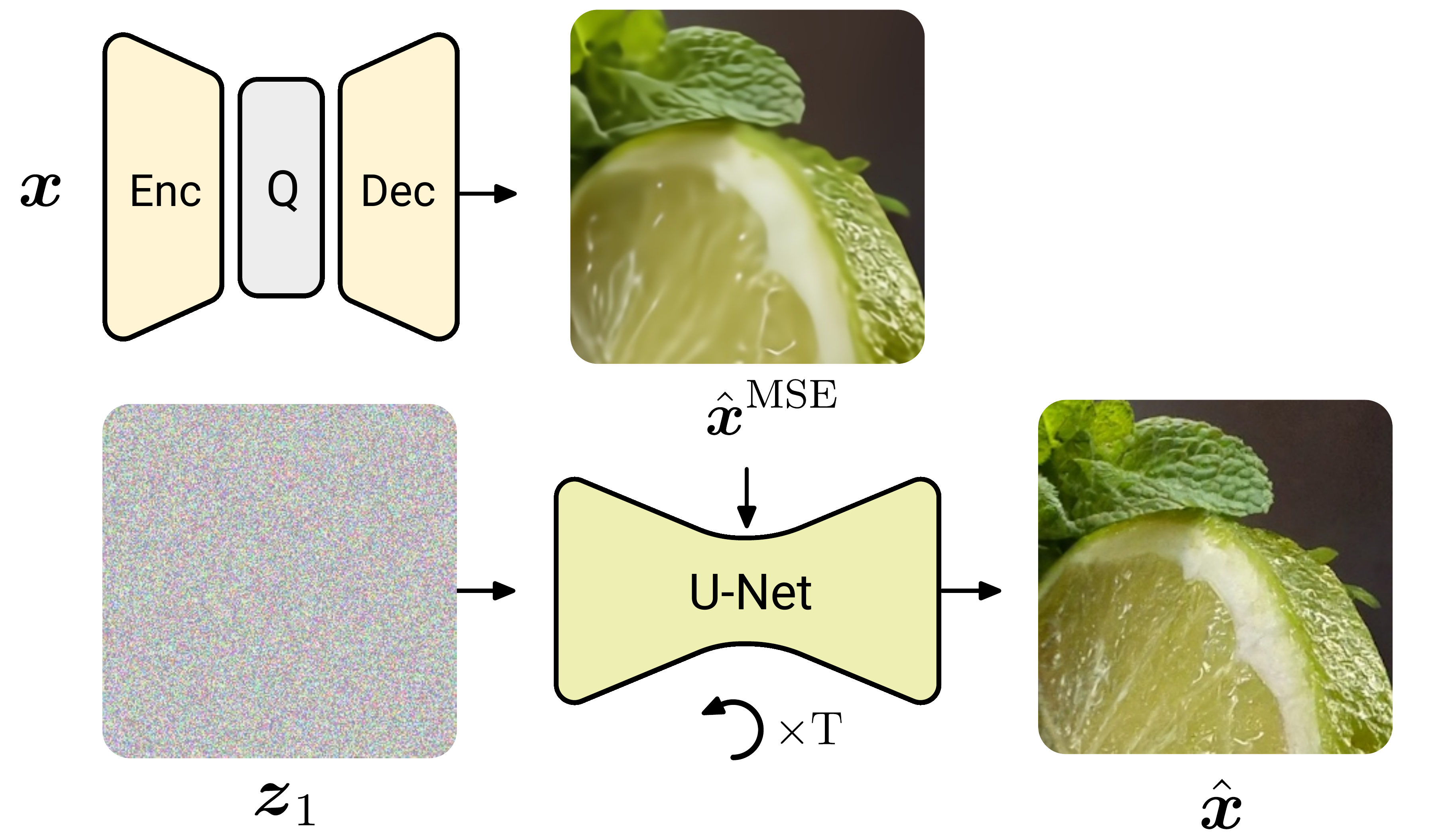}
    \caption{Overview of our high-fidelity diffusion (HFD) approach. The output of a standard MSE autoencoder is used by a denoising diffusion model to produce realistic samples by iteratively denoising for $\mathrm{T}$ steps.} \vspace{-.2cm}
    \label{fig:overview}
\end{figure}

\subsection{Diffusion}
\label{sec:bg_diffusion}

Diffusion models \cite{sohl2015diff,ho2020ddpm} define a process that gradually destroys the signal, typically with Gaussian noise. It is convenient to express the process in terms of marginal distributions conditioned on the original example $\vx$:
\begin{equation}
    q(\vz_t | \vx) = \mathcal{N}(\vz_t | \alpha_t \vx, \sigma_t^2 \mathbf{I}),
\end{equation}
where $\alpha_t$ decreases and $\sigma_t$ increases over time $t \in [0, 1]$. One can sample from this distribution via $\vz_t = \alpha_t \vx + \sigma_t \bm{\epsilon}_t$ 
where $\bm{\epsilon}_t \sim \mathcal{N}(0, \mathbf{I})$ is standard Gaussian noise. A generative denoising process can be learned by minimizing
\begin{equation}
    L = \mathbb{E}_{t \sim \mathcal{U}(0, 1)}\mathbb{E}_{\vz_t \sim q(\vz_t | \vx)} \Big{[} w(t) || \bm{\epsilon}_t - \hat{\bm{\epsilon}}_t||^2 \Big{]}
\end{equation}
where for a particular weighting $w(t)$ \cite{kingma2021vdm}, $L$ corresponds to a negative variational lower bound on $\log p(\vx)$, although in practice a constant weighting $w(t) = 1$ has been found superior for image quality \cite{ho2020ddpm}. Here, $\hat{\bm{\epsilon}}_t$ can be the prediction of a neural network $f(\vz_t, t, \hat{\vx}^{\mathrm{MSE}})$ which takes in the current noisy $\vz_t$ and diffusion time $t$, as well as possibly an additional context. In this paper it will be the output of a neural compression decoder,
\begin{equation}
\hat{\vx}^{\mathrm{MSE}} = D(Q(E(\textbf x))),
\label{eq:autoenc}
\end{equation}
where $E, D$ represents an autoencoder trained for MSE, and Q is a quantizer. We use ELIC~\cite{he2022elic} for $E, D$, see Sec.~\ref{sec:autoenc}.

Moreover, instead of learning $\hat{\bm{\epsilon}}_t$ directly, in this paper \textit{v} prediction is used, which is more stable towards $t \to 1$ \cite{salimans2022progressive} and has been used in high resolution tasks \cite{ho2022imagenvideo}. In short, the neural net predicts $\hat{\vv}_t = f(\vz_t, t, \hat{\vx}^{\mathrm{MSE}})$ which can be converted using $\hat{\bm{\epsilon}}_t = \sigma_t \vz_t + \alpha_t \hat{\vv}_t$. Intuitively, \textit{v} prediction is approximately \textit{x} prediction when $t \to 1$ (whereas the sampling with $\epsilon$ prediction could be numerically unstable) while it is approximately $\epsilon$ prediction near $t \to 0$ (where $x$ prediction would result in inferior sample quality).
To draw samples from the diffusion model, one defines a grid of timesteps $1, 1 - 1/T, 1 - 2/T, \ldots, 1/T$, and runs the denoising process from Gaussian noise $\vz_T \sim \mathcal{N}(0, \mathbf{I})$ after which $\vz_t$ is iteratively updated with $p(\vz_{t - 1/T} | \vz_t)$. For more details see Appendix~A.

\subsection{Rectified flow}
\label{bg:rectified_flow}
Another closely related approach called rectified flow~\cite{liu2022flowsstraight} aims to find a mapping between two arbitrary marginal distributions. Assuming we want to map some data distribution $p(\vx)$ to some other arbitrary distribution $p(\vz)$, we first define a (possibly random) pairing between samples from these distributions.
For example, we could draw as many samples from a standard normal distribution as needed to match the size of the data points $\vx_1, \vx_2, \ldots$ and create a pairing $(\vx_1, \vz_1), (\vx_2, \vz_2), \ldots$ between which a flow is learned via:
\begin{equation}
    L_i =  \mathbb{E}_{t \sim \mathcal{U}(0, 1)} \Big{[} || \vv_i - f(t \vx_i + (1 - t) \vz_i) ||^2 \Big{]},
\end{equation}
where $\vv_i = \vx_i - \vz_i$. After training the model, one can improve the pairing given the flow model and train the model again. 

\section{Method}
\begin{figure}[t]
    \centering
    \input{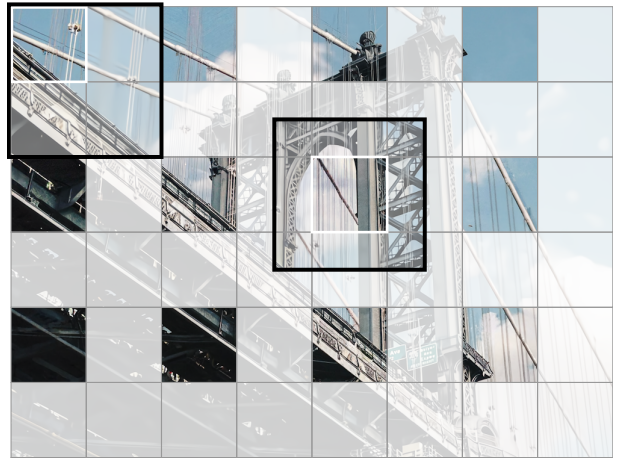}
    \caption{Our score-based models are trained on $256 \times 256$ pixel image patches. To handle arbitrary resolutions, we apply models patch-wise. We first generate a full patch while conditioning on any pixels of the input image and any already reconstructed pixels within the window (black square). We then copy the central $128 \times 128$ pixels (white square) into the final reconstruction and discard the border, which is only used to condition the model. Near the image border, context pixels are shifted relative to the central pixels (top left). By dividing patches into 4 groups, batches of image patches can easily be generated in parallel.}
    \label{fig:outpainting}
\end{figure}

On a high level, our approach consists of two components (see Fig.~\ref{fig:overview}):
first, we use a standard CNN-based autoencoder $E, D$ trained for MSE to store a lossy version of the input image to disk (detailed in Sec.~\ref{sec:autoenc}). Then, we apply a diffusion process to recover and add detail discarded by the autoencoder.
The bit-rate to encode a given image is entirely determined by $E$, since the diffusion process does not require additional bits. This two-step approach can be theoretically justified as follows. The second step approximates sampling from the posterior distribution over images $\vx$ given the output $\hat{\vx}^{\mathrm{MSE}}$ (Eq.~\ref{eq:autoenc}) of the autoencoder, $\hat{\vx} \sim p(\vx \mid \hat{\vx}^{\mathrm{MSE}})$. The MSE of this reconstruction is upper-bounded by twice the MSE of the first stage \cite{blau2018tradeoff}
\begin{align}
    \mathbb{E}[\|\hat{\vx} - \vx\|^2]
    &= 2 \, \mathbb{E}[\|\mathbb{E}[\hat{\vx} \mid \hat{\vx}^{\mathrm{MSE}}] - \vx\|^2] \\
    &= 2 \, \mathbb{E}[\|\mathbb{E}[\vx \mid \hat{\vx}^{\mathrm{MSE}}] - \vx\|^2] \\
    &\leq 2 \, \mathbb{E}[\|\hat{\vx}^{\mathrm{MSE}} - \vx\|^2],
\end{align}
with equality when the first-stage decoder is optimal. That is, given enough representational power, the loss optimized in the first stage also minimizes the distortion of the final reconstruction, and a lack of end-to-end training poses no theoretical limitation to the model's performance. The only way to further improve upon the theoretical performance of this approach (i.e., reducing MSE while maintaining perfect realism) would require a random coding approach with a shared source of randomness \cite{theis2021advantages,zhang2021universal}. However, these approaches can be expensive \cite{theis2022diff} and are currently not widely used.

\subsection{Autoencoder} \label{sec:autoenc}

The lossy MSE-optimized autoencoder is not the focus of this paper, and similar to Agustsson \etal \cite{agustsson2022mr} we use the recently proposed ELIC architecture~\cite{he2022elic} for the autoencoder (using $C=256$ channels throughout).
The quantized representation of an image produced by this autoencoder is entropy coded and written to disk. To do this, we use the channel-autoregressive entropy model proposed by Minnen~\etal~\cite{minnen2020channel}. Please see the cited work for details. We will refer to this model as ``MSE (Ours)''.

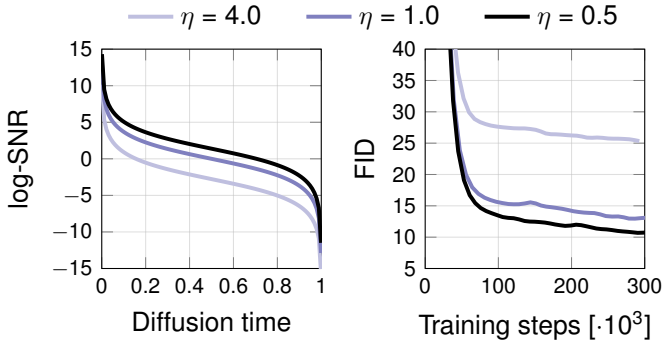
\begin{figure}[b]
    \centering
    \hspace{-1.0cm}
        \begin{tikzpicture}
        \begin{axis}[
            xshift=-4.3cm,
            width=4.5cm,
            height=4.5cm,
            xmin=0,
            xmax=1,
            ymin=-15,
            ymax=15,
            ytick={-15,-10,...,15},
            ylabel={log-SNR},
            xlabel={Diffusion time},
            grid=both,
            grid style={opacity=0.5},
        ]
        
        \addplot[blue!50!black!25!white,line width=1.5pt,domain=0.001:0.999,samples=100] 
        {-2 * (ln(tan(180 * x / 2)) + ln(4))};
        \addplot[blue!50!black!50!white,line width=1.5pt,domain=0.001:0.999,samples=100] 
        {-2 * (ln(tan(180 * x / 2)) + ln(1))};
        \addplot[black,line width=1.5pt,domain=0.001:0.999,samples=100]
        {-2 * (ln(tan(180 * x / 2)) + ln(0.5))};
        
        \end{axis}
        \begin{axis}[
            width=4.5cm,
            height=4.5cm,
            ymin=5,
            ymax=40,
            xmin=0,
            xmax=300000,
            ylabel={FID},
            xlabel={Training steps [$\cdot 10^3$]},
            ytick={5,10,...,50},
            grid=both,
            grid style={opacity=0.5},
            scaled x ticks=base 10:-3,
            xtick scale label code/.code={},
            legend style={
              draw=none,
              /tikz/every even column/.append style={column sep=0.5cm},
              at={(-0.23,1.25)},
              anchor=north},
            legend columns={-1},
        ]
            \addplot[blue!50!black!25!white,line width=1.5pt] table[x=step,y=fid,col sep=comma] {figures/noise_schedule/schedule_ds=4.0.csv};
            \addlegendentry{$\eta = 4.0$};
            
            \addplot[blue!50!black!50!white,line width=1.5pt] table[x=step,y=fid,col sep=comma] {figures/noise_schedule/schedule_ds=1.0.csv};
            \addlegendentry{$\eta = 1.0$};
            
            \addplot[black,line width=1.5pt] table[x=step,y=fid,col sep=comma] {figures/noise_schedule/schedule_ds=0.5.csv};
            \addlegendentry{$\eta = 0.5$};
        \end{axis}
    \end{tikzpicture}
    \caption{
    \textit{Left:}
    The shifted schedule which focuses more on details as used in HFD. Note that schedule is shifted in the \textit{opposite} direction of \cite{chen2023importance,hoogeboom2023simple}, as it focuses on detail opposed to global structure.
    \textit{Right:}
    FID as a function of training time for three different noise schedules. Changing the noise schedule so that fewer steps are spent processing noisy images improved performance ($\eta = 0.5$). This is in contrast to text-to-image models, where noisier schedules were found to perform better ($\eta = 4.0$) \cite{hoogeboom2023simple}. Here, FID was calculated using a validation set of 50k examples and 10k samples from the model.}
    \label{fig:fid_noise_schedule}
\end{figure}

\subsection{Score-based decoder models}
Given the autoencoder reconstruction $\hat{\vx}^{\mathrm{MSE}}$, we explore two approaches to produce a more realistic version, based on either diffusion models or rectified flows. These generate the final reconstruction by iteratively sampling from the respective generative process (Sections~\ref{sec:bg_diffusion},~\ref{bg:rectified_flow}) as follows.

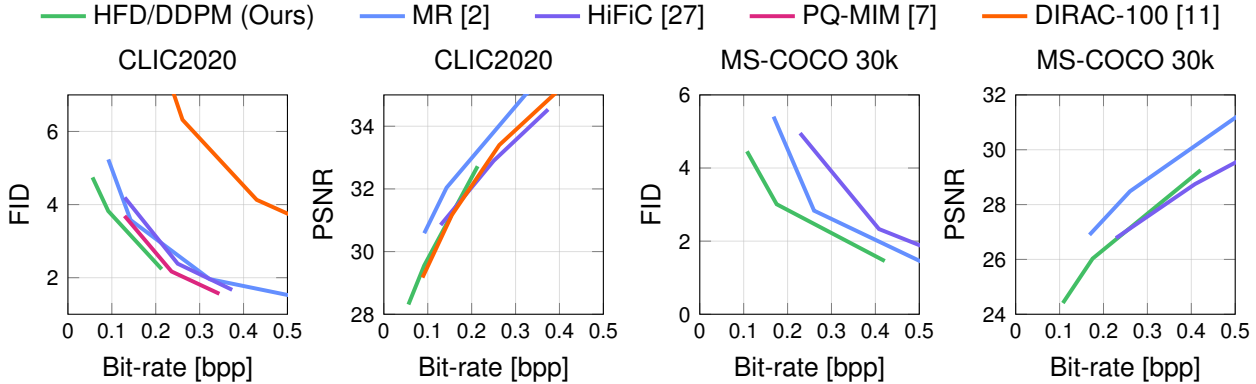
\begin{figure*}[t]
    \centering
        \begin{tikzpicture}
        
        \begin{axis}[
            title={CLIC2020},
            width=4.5cm,
            height=4.5cm,
            ymin=1,
            ymax=7,
            xmin=0,
            xmax=0.5,
            xlabel={Bit-rate [bpp]},
            ylabel={FID},
            xtick={0,0.1,...,0.6},
            grid=both,
            grid style={opacity=0.5},
            legend pos=outer north east,
            legend style={
                at={(-0.5cm,4.7cm)},
                draw=none,
                legend columns=5,
                /tikz/every even column/.append style={column sep=0.5cm}},
            legend cell align=left,
        ]
            
            \addplot[colorours,line width=1.5pt] table[x=hfd_ddpm_bpp,y=hfd_ddpm_fid,col sep=comma] {figures/metrics/fid_psnr_clic2020.txt};
            \addlegendentry{HFD/DDPM (Ours)};
            
            
            
            \addplot[colormr,line width=1.5pt] table[x=mr_bpp,y=mr_fid,col sep=comma] {figures/metrics/fid_psnr_clic2020.txt};
            \addlegendentry{MR \cite{agustsson2022mr}};
            
            \addplot[colorhific,line width=1.5pt] table[x=hific_bpp,y=hific_fid,col sep=comma] {figures/metrics/fid_psnr_clic2020.txt};
            \addlegendentry{HiFiC \cite{mentzer2020hific}};
            
            \addplot[colorpq,line width=1.5pt] table[x=meta_pq_bpp,y=meta_pq_fid,col sep=comma] {figures/metrics/fid_psnr_clic2020.txt};
            \addlegendentry{PQ-MIM \cite{el2022image}};
            
            \addplot[colorqc,line width=1.5pt] table[x=qualcom_dirac_100_bpp,y=qualcom_dirac_100_fid,col sep=comma] {figures/metrics/fid_psnr_clic2020.txt};
            \addlegendentry{DIRAC-100 \cite{goose2023neural}};
            
        \end{axis}
        
        \begin{axis}[
            xshift=4.2cm,
            title={CLIC2020},
            width=4.5cm,
            height=4.5cm,
            ymin=28,
            ymax=35,
            xmin=0,
            xmax=0.5,
            xlabel={Bit-rate [bpp]},
            ylabel={PSNR},
            xtick={0,0.1,...,0.6},
            grid=both,
            grid style={opacity=0.5},
            legend pos=outer north east,
            legend style={draw=none},
            legend cell align=left,
        ]
            \addplot[colorours,line width=1.5pt] table[x=hfd_ddpm_bpp,y=hfd_ddpm_psnr,col sep=comma] {figures/metrics/fid_psnr_clic2020.txt};
            
            
            
            \addplot[colormr,line width=1.5pt] table[x=mr_bpp,y=mr_psnr,col sep=comma] {figures/metrics/fid_psnr_clic2020.txt};
            
            \addplot[colorhific,line width=1.5pt] table[x=hific_bpp,y=hific_psnr,col sep=comma] {figures/metrics/fid_psnr_clic2020.txt}; 
            
            \addplot[colorqc,line width=1.5pt] table[x=qualcom_dirac_100_bpp,y=qualcom_dirac_100_psnr,col sep=comma] {figures/metrics/fid_psnr_clic2020.txt}; 
        \end{axis}

        \begin{axis}[
            title={MS-COCO 30k},
            xshift=8.4cm,
            width=4.5cm,
            height=4.5cm,
            ymin=0,
            ymax=6,
            xmin=0,
            xmax=0.5,
            ylabel={FID},
            xlabel={Bit-rate [bpp]},
            xtick={0,0.1,...,0.6},
            grid=both,
            grid style={opacity=0.5},
            legend pos=outer north east,
            legend style={draw=none},
            legend cell align=left,
        ]
        
            \addplot[colorours,line width=1.5pt] table[x=hfd_ddpm_bpp,y=hfd_ddpm_fid,col sep=comma] {figures/metrics/fid_psnr_coco.txt};
            
            
            
            \addplot[colormr,line width=1.5pt] table[x=mr_bpp,y=mr_fid,col sep=comma] {figures/metrics/fid_psnr_coco.txt};
            
            \addplot[colorhific,line width=1.5pt] table[x=hific_bpp,y=hific_fid,col sep=comma] {figures/metrics/fid_psnr_coco.txt};
            
        \end{axis}
        
        \begin{axis}[
            title={MS-COCO 30k},
            xshift=12.6cm,
            width=4.5cm,
            height=4.5cm,
            ymin=24,
            ymax=32,
            xmin=0,
            xmax=0.5,
            ylabel={PSNR},
            xlabel={Bit-rate [bpp]},
            xtick={0,0.1,...,0.6},
            grid=both,
            grid style={opacity=0.5},
            legend pos=outer north east,
            legend style={draw=none},
            legend cell align=left,
        ]
        
            \addplot[colorours,line width=1.5pt] table[x=hfd_ddpm_bpp,y=hfd_ddpm_psnr,col sep=comma] {figures/metrics/fid_psnr_coco.txt};
            
            
            
            \addplot[colormr,line width=1.5pt] table[x=mr_bpp,y=mr_psnr,col sep=comma] {figures/metrics/fid_psnr_coco.txt};
            
            \addplot[colorhific,line width=1.5pt] table[x=hific_bpp,y=hific_psnr,col sep=comma] {figures/metrics/fid_psnr_coco.txt};
        
        \end{axis}
    \end{tikzpicture}
    \caption{Realism and distortion as measured by FID and PSNR for various methods evaluated on MS-COCO 30k and CLIC20. HFD/DDPM is able to generate \textit{realistic images at extremely low bit-rates}, surpassing all existing methods in terms of rate-FID curves. }
    \label{fig:metrics}
\end{figure*}

\paragraph{Diffusion model}
An important property of a diffusion model is its noise schedule. It determines how quickly information is destroyed and how much of computation is spend on the generation of coarse or fine details of an image. A convenient way to express the diffusion parameters $\alpha_t, \sigma_t$ is by defining schedules in their signal-to-noise ratio (SNR = $\alpha_t^2 / \sigma_t^2$), or rather in their log-SNR schedule \cite{kingma2021vdm}. Under a variance preserving assumption (a particular flavour of diffusion models where $\alpha_t^2 = 1 - \sigma_t^2$), given the log SNR one can simply retrieve $\alpha_t^2 = \mathrm{sigmoid}(\log \mathrm{SNR}(t))$ and $\sigma_t^2 = \mathrm{sigmoid}(-\log \mathrm{SNR}(t))$. In contrast to previous work \cite{hoogeboom2023simple,chen2023importance} which found it helpful to shift the schedule towards increased levels of noise, for compression we find it beneficial to shift the schedule in the \textit{opposite} direction. Intuitively speaking, the output of the MSE-trained decoder $\hat{\vx}^{\mathrm{MSE}}$ already provides a lot of global information about the image structure. It would therefore be wasteful to dedicate a large part of the diffusion process to generation of the global structure, which is associated with high noise levels.
By shifting the schedule to use less noise, the diffusion model instead focuses on the finer details of an image. Recall that under a variance preserving process, the $\alpha$-cosine schedule is described by $-2 \log \tan (\pi t / 2)$ in log SNR, using that $\cos / \sin = 1 / \tan$ and $\cos^2(t) + \sin^2(t) = 1$. We adapt this schedule to:
\begin{equation}
\log \mathrm{SNR}(t) = - 2 \big{(}\log \tan (\pi t / 2) + \log \eta\big{)},
\end{equation}
which is shifted by $-2 \log \eta$ to reduce the amount of noise (we use $\eta$ $=$ $0.5$, see Fig.~\ref{fig:fid_noise_schedule}).
As is standard practice, the boundary effects (where log SNR tends to $\pm \infty$) at $t = 0$ and $t = 1$ are mitigated by bounding the log SNR, in this case to $\pm 15$. Combining everything, the objective can be summarized to be:
\begin{equation}
    L = \mathbb{E}_{t \sim \mathcal{U}(0, 1)}\mathbb{E}_{\veps_t \sim \mathcal{N}(0, \mathbf{I})} \Big{[} || \bm{\epsilon}_t - \hat{\bm{\epsilon}}_t(\vz_t, t, \hat{\vx}^{\mathrm{MSE}}) ||^2 \Big{]}
\end{equation}
where $\vz_t = \alpha_t \vx + \sigma_t \veps_t$,
$\hat{\bm{\epsilon}}_t = \sigma_t \vz_t + \alpha_t \hat{\vv}_t$ (the model uses \textit{v}-prediction which improves stability for higher resolution images),
$\alpha_t^2 = \mathrm{sigmoid}(\log \mathrm{SNR}(t))$,
$\sigma_t^2 = \mathrm{sigmoid}(-\log \mathrm{SNR}(t))$.
$\hat{\vv}_t$ is predicted by the neural network $f(\vz_t, t, \hat{\vx}^{\mathrm{MSE}})$ which is a U-Net concatenating $\vz_t$ and $\hat{\vx}^{\mathrm{MSE}}$ along the channel axis. 

\paragraph{Flow matching}
Recall that flow matching initially trains a mapping on unpaired examples. However, in this work we are able to use the pairing $(\bm x, \hat{\bm x^{\mathrm{MSE}}})$ given by the autoencoder.
This means that instead of conditionally mapping Gaussian samples to images as in diffusion, we learn to map autoencoder outputs directly to the uncompressed image using flow matching. 
We add a small amounts of uniform noise to the reconstructions $\hat{\vx}^{\mathrm{MSE}}$ and targets $\vx$ to ensure that an invertible flow between the distributions exist,
even though this did not seem to be necessary in practice.
We do not iteratively apply rectification as proposed by Liu \etal \cite{liu2022flowsstraight}, leaving us with a simple optimization objective:
\begin{equation*}
    L =  \mathbb{E}_t \Big{[} || (\vx - \hat{\vx}^{\mathrm{MSE}}) - f(t \vx + (1 - t) \hat{\vx}^{\mathrm{MSE}}) ||^2 \Big{]},
\end{equation*}
Instead of sampling $t$ uniformly, we found it beneficial to use $t = 1 - u^2$ where $u$ is sampled uniformly between $0$ and $1$. However, we did not extensively explore the schedule for rectified flow.

\subsection{Generation and sampling}

\paragraph{Parallelized sampling of patches}
Compression models are typically trained using fully convolutional architectures on patches, to be applied to full resolution images at test time. However, diffusion models often rely on self-attention layers which are not equivariant and whose computational complexity grows more quickly in the input dimensions.

We therefore opt to generate high-resolution images in a patchwise manner. Fortunately, in- and out-painting is relatively easy in diffusion models. In each step of the generative denoising process, already observed pixels are simply replaced by the known values corrupted by an appropriate amount of noise. While we could generate patches one-by-one, with some overlap to previous patches, this leads to low utilization of modern accelerators.

Instead, in this paper patches are divided in groups of four as visualized in Fig.~\ref{fig:outpainting}. Each group can be generated independently where each patch is a single example in the batch. Then, the next group of patches can be generated resulting in four distinct generation stages.
As is typical in diffusion, the input for the model is the current noisy state $\vz_t$ together with already previously generated parts of the patch $\hat{\vx}$, controlled by a mask $\vm$ so that the input is:
\begin{equation}
\vm \vz_t + (1 - \vm) (\alpha_t \hat{\vx}  + \sigma_t \veps_t),
\end{equation}
where $\vm$ is one for pixel locations that still need to be generated and diffusion noise is injected with $\veps_t \sim \mathcal{N}(0, \mathbf{I})$. Note here that $\hat{\vx}$ is the output of the diffusion model from previously generated patches.

This approach often works well despite only approximating proper probabilistic conditioning on the available information. Nevertheless, we find that it occasionally leads to artefacts. To overcome this issue, we only partially run the diffusion process, generating a patch of noisy pixels (Appendix~F). The next patch is then partially generated conditioned on observed noisy pixels. We then revisit patches to continue the reverse diffusion process. We find that dividing the diffusion process into 6 groups works well to eliminate any remaininig artefacts.

\begin{table}[b]
    \centering
    \caption{ \label{tab:architecture} HFD U-Net architecture\vspace{2ex}}
    \begin{tabular}{l l l l l l l}
    \toprule
        Level & $256\times$ & $128\times$ & $64\times$ & $32\times$ & $16\times$ &\\ \midrule
        Channels & 128 & 128 & 256 & 256 & 1024 \\
        Blocks & 2 & 2 & 2 & 2 & 16 \\ 
        Attention & - & - & - & - & \checkmark \\ \bottomrule 
    \end{tabular}
\end{table}

\paragraph{Noise level during sampling}
An important and sometimes forgotten hyperparameter of diffusion models is the noise level of the denoising process, which can be any value $\sqrt( \sigma_{ts}^{2\gamma} \sigma_{t \to s}^{2(\gamma - 1)})$ for $\gamma \in [0, 1]$. Here $\sigma_{st}^2$ is the diffusion variance and $\sigma_{t \to s}^{2}$ is the \textit{true} denoising variance, when conditioned on a single example (detailed in Appendix~A). For smaller noise levels ($\gamma \approx 0.0$) and larger number of denoising steps, generations tend to become blurrier. For larger noise levels ($\gamma \approx 1.0$) and smaller number of denoising steps, generations tend to become grainy and noisy. To limit the cost of sampling, we consider sampling steps from and below $250$. In this setting, we find that smaller noise levels are preferred ($\gamma = 0.0$ for MS-COCO and $\gamma = 0.1$ for CLIC20).

\subsection{Architecture}
Diffusion models generally use U-Nets \cite{ronneberger2015unet} with residual convolutional blocks and self-attention. Because convolutional layers at high resolutions are very expensive in terms of memory and computation, we limit the size of these layers as much as possible. The exact details are given in Table~\ref{tab:architecture}. The autoencoder output $\vx^{\mathrm{MSE}}$ is concatenated to the current diffusion state $\vz_t$ as the first step of the architecture. Following recent advances in diffusion \cite{saharia2022imagen,jabri2022scalable,hoogeboom2023simple}, the bulk of the computation is moved from high resolution to the lower resolution feature maps.

\section{Experiments}

\begin{figure}
    \centering
        \begin{tikzpicture}
            
        
            
        
        \begin{axis}[
            xshift=8.4cm,
            title={},
            width=4.5cm,
            height=4.5cm,
            ymin=0,
            ymax=50,
            xmin=0,
            xmax=250,
            xlabel={Steps},
            xtick={0,50,...,300},
            ylabel={FID},
            grid=both,
            grid style={opacity=0.5},
            legend pos=outer north east,
            legend style={draw=none},
            legend cell align=left,
        ]
            \addplot[color1,line width=1.5pt] table[x=steps,y=FID/256,col sep=comma] {figures/metrics/clic2020_hfd_ddpm_steps.csv};
            
            \addplot[color3,line width=1.5pt] table[x=steps,y=FID/256,col sep=comma] {figures/metrics/clic2020_rf_steps.csv};
        \end{axis}
        
        \begin{axis}[
            xshift=12.6cm,
            title={},
            width=4.5cm,
            height=4.5cm,
            ymin=25,
            ymax=35,
            xmin=0,
            xmax=250,
            xlabel={Steps},
            xtick={0,50,...,300},
            ylabel={PSNR},
            grid=both,
            grid style={opacity=0.5},
            legend style={
              draw=none,
              /tikz/every even column/.append style={column sep=0.5cm},
              at={(-0.23,1.25)},
              anchor=north},
            legend columns={-1},
        ]
            \addplot[color1,line width=1.5pt] table[x=steps,y=PSNR/RGB,col sep=comma] {figures/metrics/clic2020_hfd_ddpm_steps.csv};
            \addlegendentry{\textbf{HFD} (Ours)};
            
            \addplot[color3,line width=1.5pt] table[x=steps,y=PSNR/RGB,col sep=comma] {figures/metrics/clic2020_rf_steps.csv};
            \addlegendentry{\textbf{RF} (Ours)};
        \end{axis}
    \end{tikzpicture}
    \caption{FID and PSNR as a function of the number of steps used to simulate the SDE/ODE underlying each model on CLIC20.}
    \label{fig:fid_psnr_steps}
\end{figure}
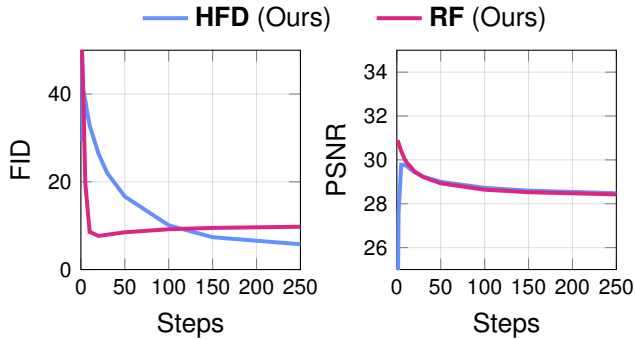


\subsection{Metrics}
We focus on the well-established metrics FID~\cite{heusel2017fid} and PSNR to measure realism and distortion, respectively. In line with previous work~\cite{mentzer2020hific,agustsson2022mr,el2022image}, we evaluate FID on patches of $256{\times}256$ pixels, see Appendix A.7 of Mentzer \etal~\cite{mentzer2020hific}.

\subsection{Datasets}
We compare on the following datasets: \textbf{Kodak}~\cite{kodak}, containing 24 images, each either $512{\times}768$px or the inverse.
From the CLIC compression challenge~\cite{clic2022},
we use the full dataset of \textbf{CLIC20}\footnote{\url{www.tensorflow.org/datasets/catalog/clic}}, which contains 428 images of varying resolutions, up to 2000px wide,
and the test set of \textbf{CLIC22}~\cite{clic2022},
which contains 30 high-resolution images, resized such that the longer side is 2048px.
While we can evaluate FID in the patched manner mentioned above on CLIC20, the other datasets are too small.
Inspired by the image generation literature (\eg,~\cite{yu2022scaling}),
recent work by Agustsson~\etal~\cite{agustsson2022mr} additionally evaluates on \textbf{MS-COCO 30k}, which we also use. We follow the preparation scheme linked in~\cite{agustsson2022mr} and compare to their published results.
It is a dataset of 30\,000 images of $256{\times}256$px each, and hence our patched FID corresponds to full FID.

\begin{figure}[t]
    \input{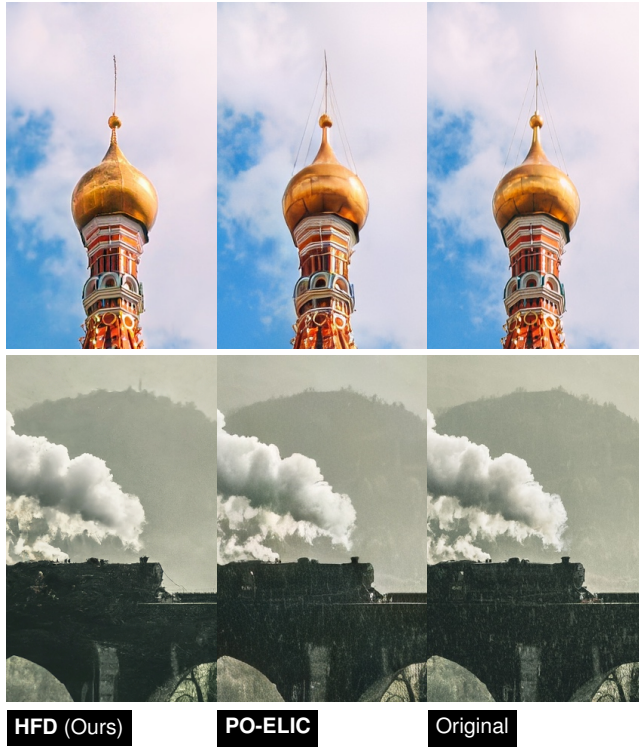}
    \caption{\textit{Failure cases}. HFD has been optimized to produce realistic images from an MSE based decoder. Consequently, high and mid frequency details can sometimes be lost or generated differently. For example, the cable lines have disappeared in the generation from HFD. Comparison at a comparable low bit-rate setting.}
    \label{fig:realism_vs_acc}
    \vspace{-.3cm}
\end{figure}

\begin{figure*}[t]
    \centering
    \input{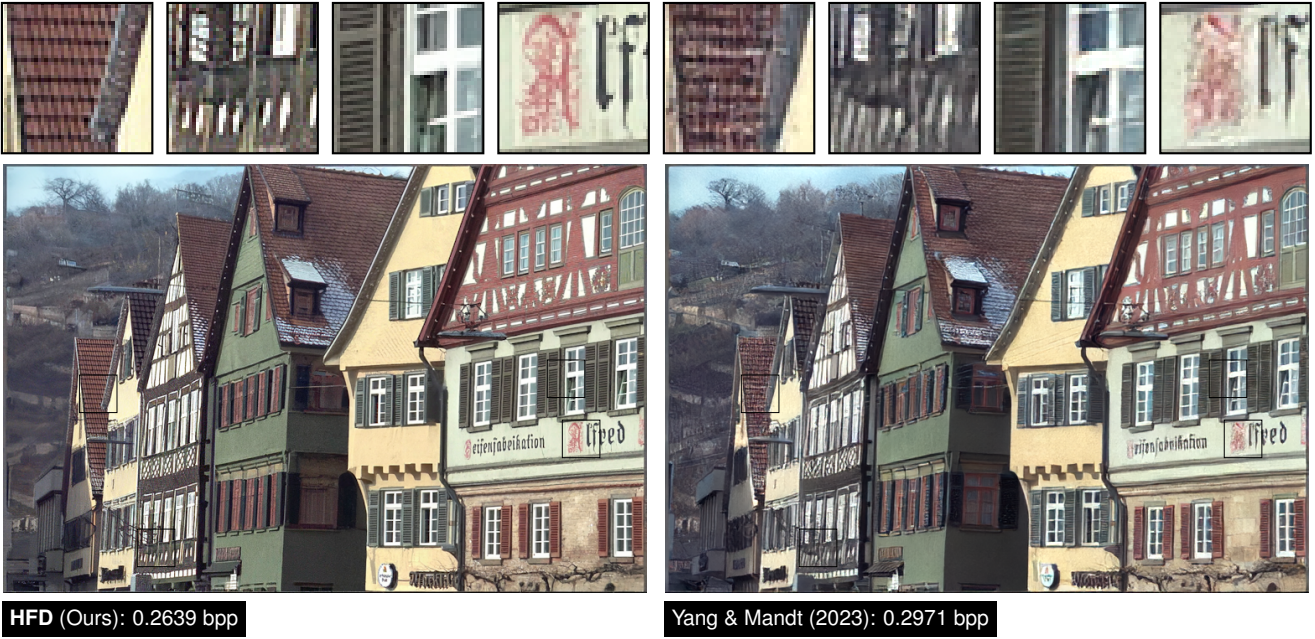}
    \caption{Qualitative comparison with the approach of Yang and Mandt \cite{yang2022diff}. We find that our model generally produces fewer artefacts at the same or lower bit-rate, despite not being trained end-to-end. Additional examples are provided in Appendix~B.}
    \label{fig:yang_mandt}
\end{figure*}

\subsection{Training}
We train our models for 2M iterations with a batch size of 256 on crops with resolution 256px. 
To get crops for training, we extract a collection of 640px crops from Internet images and encode/decode them with our MSE model. We then discard a 64px border to get pairs of 512px reconstructions and originals. This is to ensure that potential border artefacts from the MSE model are not over-represented in the training data (compared to the high-resolution evaluation set). In addition, the subset which contains people of the train partition of the MS-COCO \cite{lin2014mscoco} dataset is used. We found that the inclusion of the latter was also important to improve performance on the MS-COCO eval benchmark, 4.46 versus 6.79 in terms of FID for HFD at the low bit-rate setting. This importance may be caused by the difference in distribution between  patches from high resolution data and center crop images that are typically used for MS-COCO. 

The optimization for the model uses Adam with $\beta_1=0.9$ and $\beta_2=0.99$ and a learning rate of $10^{-4}$, with a warmup of $10000$ and half life of $400000$. Finally all evaluation is done on an exponential moving average, that is computed using $0.9999$ during training.

\subsection{Baselines}

For our models, we run \textbf{HFD/DDPM} using DDPM for sampling, 
\textbf{HFD/DDIM} using DDIM, and \textbf{RF} using rectified flows.
We compare against the following baselines. Note that not all methods publish reconstructions on all datasets, and not all datasets are big enough to compute FID reliably, so we compare against some methods only visually.
From the GAN-based image compression literature, we compare against \textbf{HiFiC}~\cite{mentzer2020hific}, \textbf{MR}~\cite{agustsson2022mr},
\textbf{PQ-MIM}~\cite{el2022image},
as well as \textbf{PO-ELIC}~\cite{he2022poelic} (the latter only has reconstructions on CLIC22, so we only compare visually).
Finally, we compare against the diffusion-based approaches
\textbf{DIRAC}~\cite{goose2023neural}, which presents FID results on CLIC20 (we use the high perceptual quality model), and \textbf{CDC}~\cite{yang2022diff} (only visually, on Kodak).

\vspace{-0.5pt}
\paragraph{Results}
As is shown in Fig.~\ref{fig:metrics}, HFD outperforms all other baselines in terms of rate-FID curves on both CLIC20 and MS-COCO 30K. On the other hand, that realism comes at the cost of distortion in terms of PSNR where other models are either better or competitive. Interestingly, FID score is improved by our proposed shifted schedule for more detail in Fig.~\ref{fig:fid_noise_schedule}, whereas the opposite direction (as proposed in the literature) worsens the performance. This confirms our hypothesis that HFD benefits more from focusing on finer details in images.

Furthermore, Fig.~\ref{fig:fid_psnr_steps} shows that rectified flows outperform HFD when the number of steps is constrained to less than approximately 100 steps. However, in line with results in the literature \cite{ho2020ddpm,liu2022flowsstraight} HFD outperforms the rectified flow for a larger sampling budget. In terms of distortion, larger sampling budgets typically result in lower PSNR. 
Qualitative comparisons can be found in Figs.~\ref{fig:hero}, \ref{fig:text-to-image}, \ref{fig:pq_mim}, \ref{fig:yang_mandt} in addition to further comparisons in the Appendix. 

\vspace{-.1cm}
\paragraph{Realism versus Distortion}
HFD can be seen as a method that favors realism over distortion. We find that this causes it to sometimes produce reconstructions which are less accurate than other methods. Example failure cases are provided in Fig.~\ref{fig:realism_vs_acc}. These images contain details that have largely vanished from the autoencoder output $\hat{\vx}^{\mathrm{MSE}}$, for example the cable lines or the grain on black surfaces. HFD also has a denoising effect causing reconstructions of noisy images to look less like the input, despite looking realistic. We find that this can be addressed by additionally encoding the absolute residuals at low resolution and very small bit-rates, and conditioning the diffusion model on this additional signal (Appendix~G).

\section{Discussion}
In this paper we have demonstrated that HFD consistently outperforms existing methods in terms of FID on multiple datasets, especially at low bit-rates. This was enabled by modifications to the diffusion approach specifically aimed at the compression setting, most importantly shifting the noise schedule. Furthermore, we show that the rectified flow outperforms diffusion with very few sampling steps although for larger numbers of steps the flow is still outperformed by its diffusion counterpart. We see several avenues for further improvement. One of the main challenges for future work will be to improve the sampling speed of diffusion-based compression approaches with techniques such as progressive distillation \cite{salimans2022distill}.

\subsection*{Acknowledgments}
The authors would like to thank Erfan Noury for providing HEVC and VVC reconstructions for the CLIC22 dataset, Ruihan Yang for providing Kodak reconstructions~\cite{yang2022diff}, David Minnen for help obtaining MSE based reconstructions used in an earlier implementation, and Ben Poole for feedback on the manuscript.

{
\small
\putbib
}

\end{bibunit}

\appendix
\onecolumn

\begin{bibunit}
      \section{Additional details on diffusion models}
    \label{app:additional_details_diffusion}
    This section contain additional details on the diffusion model. Recall that the marginal distribution of the diffusion process is defined by:
    \begin{equation}
    q(\vz_t | \vx) = \mathcal{N}(\vz_t | \alpha_t \vx, \sigma_t^2 \mathbf{I}),
\end{equation}
where $\alpha_t, \sigma_t \in [0, 1]$ and under a variance preserving process $\alpha_t^2 = 1 - \sigma_t^2$. Assuming this process is Markov, we can write the transition probability as:
\begin{equation}
    q(\vz_t | \vz_{t-1}) = \mathcal{N}(\vz_t | \vz_{s} \alpha_{ts}, \sigma_{ts}^2 \mathbf{I})
\end{equation}
where $s < t$, $\alpha_{ts} = \alpha_t / \alpha_s$ and $\sigma_{ts}^2 = \sigma_t^2 - \alpha_{t|s}^2 \sigma_s^2$. Using the two equations above and that the process is Markov, one can derive the denoising posterior conditioned on a single example $\vx$:
\begin{equation}
    q(\vz_{s} | \vz_t, \vx) = \mathcal{N}(\vz_s | \boldsymbol{\mu}_{t \to s}(\vz_t, \vx), \sigma_{t \to s}^2 \mathbf{I}),
\end{equation}
where $\boldsymbol{\mu}_{t \to s} = \alpha_{ts} \frac{\sigma_s^2}{\sigma_t^2} \vz_t + \alpha_s \frac{\sigma_{ts}^2}{\sigma_t^2} \vx$ \cite{kingma2021vdm}. The optimal generative denoising process $p(\vz_s | \vz_t)$ tends to $q(\vz_{s} | \vz_t, \mathbb{E}[\vx | \vz_t])$ when $s \to t$ \cite{song2021score} which shows that it suffices to learn $\hat{\vx} = f(\vz_t, t)$ with a neural network. However, under a constrained number of steps, we find that the variance in $p(\vz_s | \vz_t)$ can make a difference in sample quality (too noisy or too blurry). Following \cite{saharia2022imagen} we use the formulation:
\begin{equation}
    p(\vz_s | \vz_t) = \mathcal{N}(\vz_s | \boldsymbol{\mu}_{t \to s}(\vz_t, \hat{\vx}), \sigma_{ts}^{2\gamma} \sigma_{t \to s}^{2(\gamma - 1)}) \mathbf{I})
\end{equation}
where $\gamma \in [0, 1]$ is a hyperparameter that interpolates (in log-space) between the noise of the diffusion transition variance $\sigma_{ts}^2$ and true denoising variance (for a single example) $\sigma_{t \to s}^{2}$. As a rule-of-thumb, for smaller number of sampling steps, $\gamma$ should be smaller. Note that this setting only influences the DDPM (sometimes referred to as ancestral) sampler \cite{ho2020ddpm}, the DDIM \cite{song2021ddim} sampler does not use this denoising variance.

\section{Further rate distortion results}
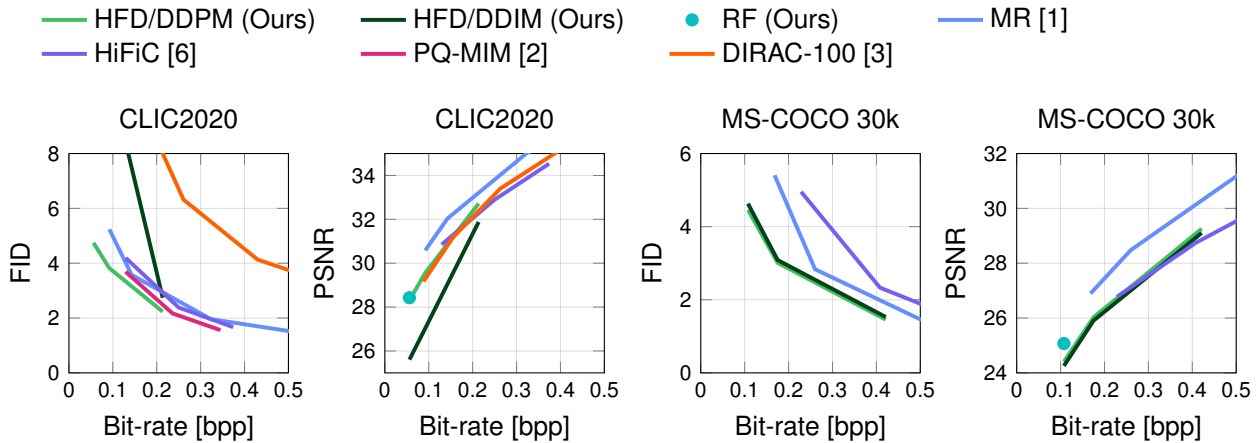
\begin{figure*}[ht]
    \centering
        \begin{tikzpicture}
        
        \begin{axis}[
            title={CLIC2020},
            width=4.5cm,
            height=4.5cm,
            ymin=0,
            ymax=8,
            xmin=0,
            xmax=0.5,
            xlabel={Bit-rate [bpp]},
            ylabel={FID},
            xtick={0,0.1,...,0.6},
            grid=both,
            grid style={opacity=0.5},
            legend pos=outer north east,
            legend style={
                at={(-0.5cm,5.0cm)},
                draw=none,
                legend columns=4,
                /tikz/every even column/.append style={column sep=0.5cm}},
            legend cell align=left,
        ]
            
            \addplot[colorours,line width=1.5pt] table[x=hfd_ddpm_bpp,y=hfd_ddpm_fid,col sep=comma] {figures/metrics/fid_psnr_clic2020.txt};
            \addlegendentry{HFD/DDPM (Ours)};
            
            \addplot[coloroursddim,line width=1.5pt] table[x=hfd_ddim_bpp,y=hfd_ddim_fid,col sep=comma] {figures/metrics/fid_psnr_clic2020.txt};
            \addlegendentry{HFD/DDIM (Ours)};
            
            \addplot[coloroursrf,only marks] table[x=rf_bpp,y=rf_fid,col sep=comma] {figures/metrics/fid_psnr_clic2020.txt};
            \addlegendentry{RF (Ours)};
            
            \addplot[colormr,line width=1.5pt] table[x=mr_bpp,y=mr_fid,col sep=comma] {figures/metrics/fid_psnr_clic2020.txt};
            \addlegendentry{MR \cite{agustsson2022mr}};
            
            \addplot[colorhific,line width=1.5pt] table[x=hific_bpp,y=hific_fid,col sep=comma] {figures/metrics/fid_psnr_clic2020.txt};
            \addlegendentry{HiFiC \cite{mentzer2020hific}};
            
            \addplot[colorpq,line width=1.5pt] table[x=meta_pq_bpp,y=meta_pq_fid,col sep=comma] {figures/metrics/fid_psnr_clic2020.txt};
            \addlegendentry{PQ-MIM \cite{el2022image}};
            
            \addplot[colorqc,line width=1.5pt] table[x=qualcom_dirac_100_bpp,y=qualcom_dirac_100_fid,col sep=comma] {figures/metrics/fid_psnr_clic2020.txt};
            \addlegendentry{DIRAC-100 \cite{goose2023neural}};
            
        \end{axis}
        
        \begin{axis}[
            xshift=4.2cm,
            title={CLIC2020},
            width=4.5cm,
            height=4.5cm,
            ymin=25,
            ymax=35,
            xmin=0,
            xmax=0.5,
            xlabel={Bit-rate [bpp]},
            ylabel={PSNR},
            xtick={0,0.1,...,0.6},
            grid=both,
            grid style={opacity=0.5},
            legend pos=outer north east,
            legend style={draw=none},
            legend cell align=left,
        ]
            \addplot[colorours,line width=1.5pt] table[x=hfd_ddpm_bpp,y=hfd_ddpm_psnr,col sep=comma] {figures/metrics/fid_psnr_clic2020.txt};
            
            \addplot[coloroursddim,line width=1.5pt] table[x=hfd_ddim_bpp,y=hfd_ddim_psnr,col sep=comma] {figures/metrics/fid_psnr_clic2020.txt};
            
            \addplot[coloroursrf,only marks] table[x=rf_bpp,y=rf_psnr,col sep=comma] {figures/metrics/fid_psnr_clic2020.txt};
            
            \addplot[colormr,line width=1.5pt] table[x=mr_bpp,y=mr_psnr,col sep=comma] {figures/metrics/fid_psnr_clic2020.txt};
            
            \addplot[colorhific,line width=1.5pt] table[x=hific_bpp,y=hific_psnr,col sep=comma] {figures/metrics/fid_psnr_clic2020.txt}; 
            
            \addplot[colorqc,line width=1.5pt] table[x=qualcom_dirac_100_bpp,y=qualcom_dirac_100_psnr,col sep=comma] {figures/metrics/fid_psnr_clic2020.txt}; 
        \end{axis}

        \begin{axis}[
            title={MS-COCO 30k},
            xshift=8.4cm,
            width=4.5cm,
            height=4.5cm,
            ymin=0,
            ymax=6,
            xmin=0,
            xmax=0.5,
            ylabel={FID},
            xlabel={Bit-rate [bpp]},
            xtick={0,0.1,...,0.6},
            grid=both,
            grid style={opacity=0.5},
            legend pos=outer north east,
            legend style={draw=none},
            legend cell align=left,
        ]
        
            \addplot[colorours,line width=1.5pt] table[x=hfd_ddpm_bpp,y=hfd_ddpm_fid,col sep=comma] {figures/metrics/fid_psnr_coco.txt};
            
            \addplot[coloroursddim,line width=1.5pt] table[x=hfd_ddim_bpp,y=hfd_ddim_fid,col sep=comma] {figures/metrics/fid_psnr_coco.txt};
            
            \addplot[coloroursrf,only marks] table[x=rf_bpp,y=rf_fid,col sep=comma] {figures/metrics/fid_psnr_coco.txt};
            
            \addplot[colormr,line width=1.5pt] table[x=mr_bpp,y=mr_fid,col sep=comma] {figures/metrics/fid_psnr_coco.txt};
            
            \addplot[colorhific,line width=1.5pt] table[x=hific_bpp,y=hific_fid,col sep=comma] {figures/metrics/fid_psnr_coco.txt};
        \end{axis}
        
        \begin{axis}[
            title={MS-COCO 30k},
            xshift=12.6cm,
            width=4.5cm,
            height=4.5cm,
            ymin=24,
            ymax=32,
            xmin=0,
            xmax=0.5,
            ylabel={PSNR},
            xlabel={Bit-rate [bpp]},
            xtick={0,0.1,...,0.6},
            grid=both,
            grid style={opacity=0.5},
            legend pos=outer north east,
            legend style={draw=none},
            legend cell align=left,
        ]
        
            \addplot[colorours,line width=1.5pt] table[x=hfd_ddpm_bpp,y=hfd_ddpm_psnr,col sep=comma] {figures/metrics/fid_psnr_coco.txt};
            
            \addplot[coloroursddim,line width=1.5pt] table[x=hfd_ddim_bpp,y=hfd_ddim_psnr,col sep=comma] {figures/metrics/fid_psnr_coco.txt};
            
            \addplot[coloroursrf,only marks] table[x=rf_bpp,y=rf_psnr,col sep=comma] {figures/metrics/fid_psnr_coco.txt};
            
            \addplot[colormr,line width=1.5pt] table[x=mr_bpp,y=mr_psnr,col sep=comma] {figures/metrics/fid_psnr_coco.txt};
            
            \addplot[colorhific,line width=1.5pt] table[x=hific_bpp,y=hific_psnr,col sep=comma] {figures/metrics/fid_psnr_coco.txt};
        
        \end{axis}
    \end{tikzpicture}
    \caption{Realism and distortion as measured by FID and PSNR for various methods evaluated on MS-COCO 30k and CLIC20. HFD/DDPM is able to generate \textit{realistic images at impressively low bitrates}, surpassing all existing methods in terms of rate-FID curves. It is also worth noting that this model considerably outperforms DIRAC-100, the only other existing diffusion approach for high resolution images.}
    \label{fig:metrics_additional}
\end{figure*}

In Fig.~\ref{fig:metrics_additional}, the same comparison from the main paper is included, but now including the DDIM \cite{song2021ddim} sampler for HFD and the rectified flow result. Our HFD/DDPM is the best performing model in terms of FID score.

\clearpage

    \section{Additional results obtained with text-to-image models}
    \begin{figure}[H]
        \centering
        \input{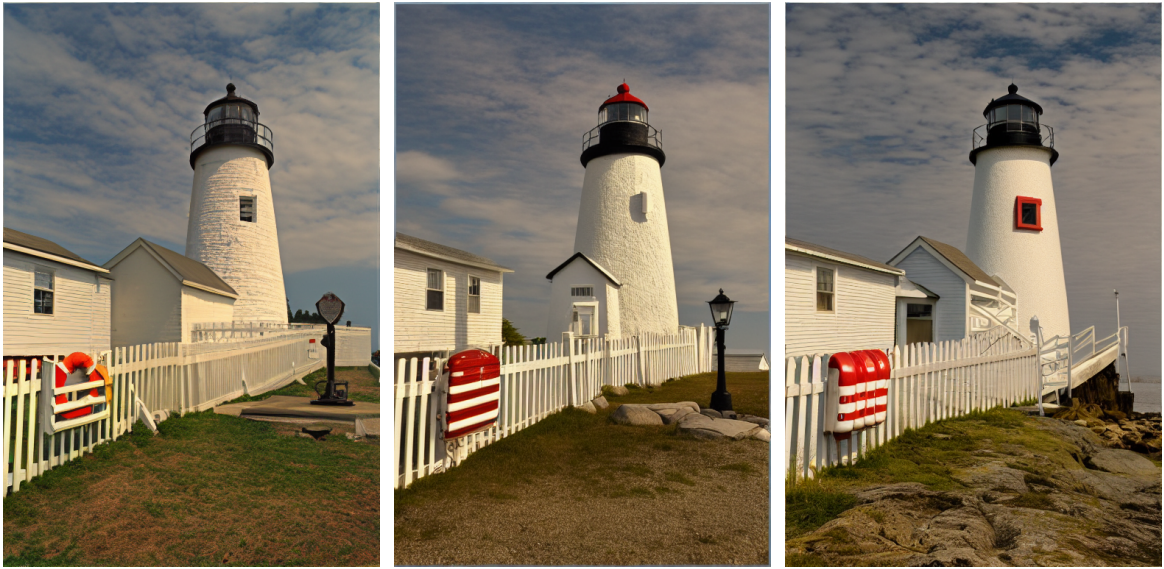}
        \caption{
            Image reconstructions obtained with Stable Diffusion \cite{rombach2022sd} by conditioning on a $4\times$ downsampled image together with the text ``A lighthouse in Maine behind a white fence with a red life buoy hanging on it.'' Depending on the choice of parameters, reconstructions are more or less faithful to the original image. However, we were unable to achieve a level of fidelity that we would deem acceptable for the task of image compression.
        }
    \end{figure}

    \begin{figure}[h!]
        \centering
        \input{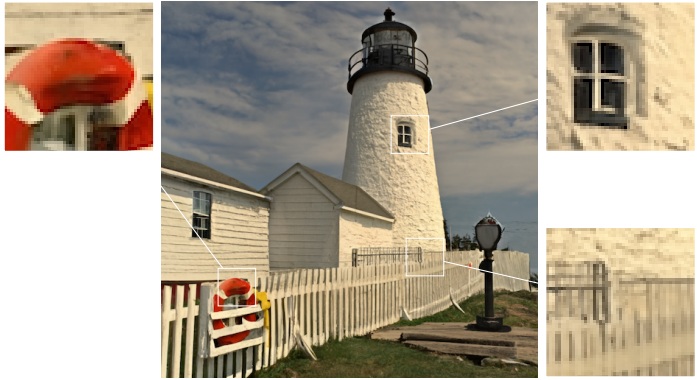}
        \caption{
            Image reconstruction of a $512 \times 512$ image obtained with Imagen \cite{saharia2022imagen} by conditioning on a $8\times$ downsampled image together with the text ``A lighthouse in Maine behind a white fence with a red life buoy hanging on it.''}
    \end{figure}

    \clearpage

    \section{Additional comparisons with Yang \& Mandt \cite{yang2022diff}}

     \begin{figure}[h!]
         \input{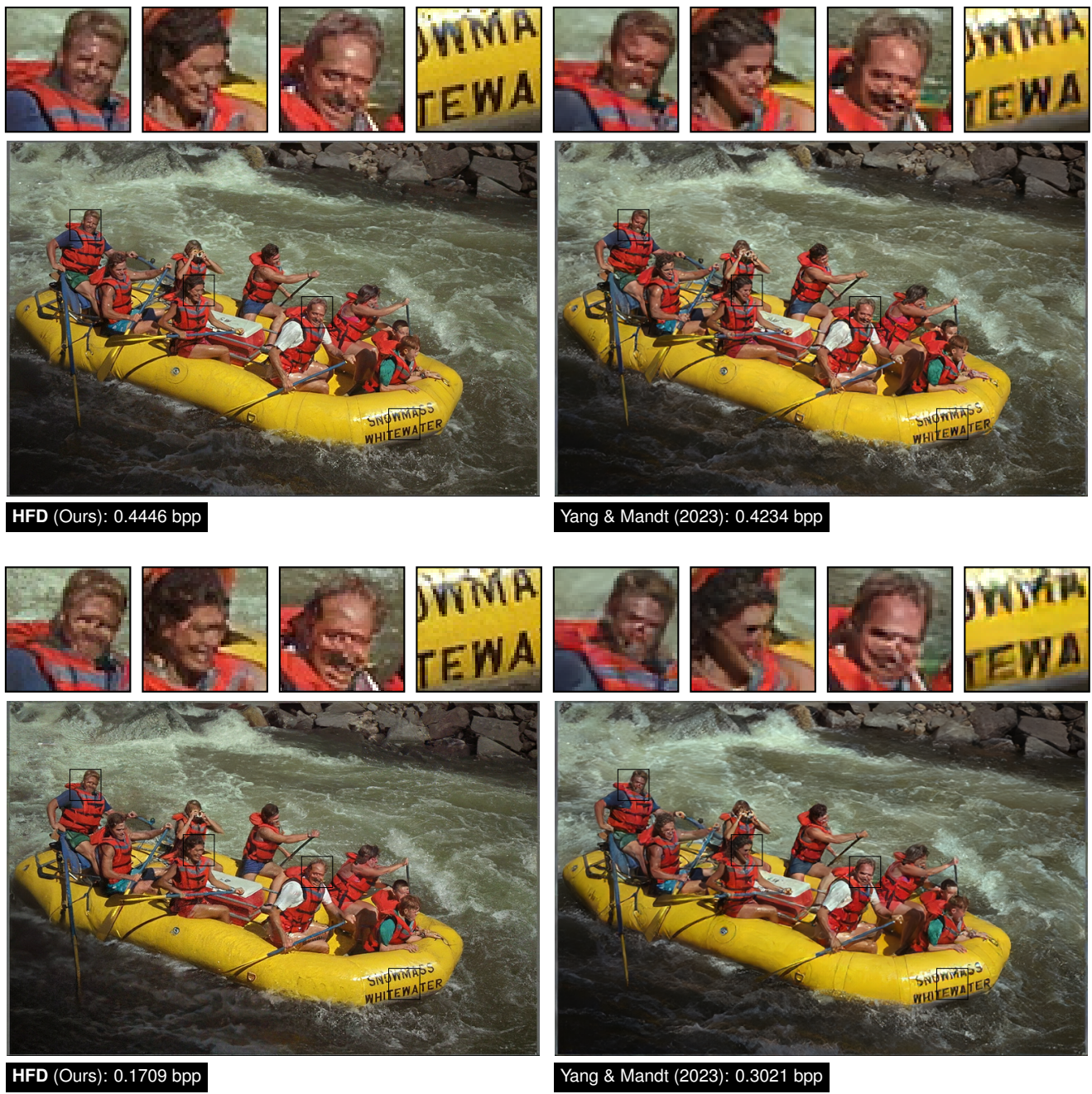}
         \vspace{0.01cm}
         \caption{Our model generally compares favorably to that of Yang \& Mandt \cite{yang2022diff} in terms of perceptual quality when evaluated at similar bit-rates (upper row) or even using a significantly lower bit-rate (lower row). Nevertheless, as for other generative compression methods, small faces remain a challenge at very low bit-rates.}
    \end{figure}

     \newpage

    \begin{figure}[H]
        \input{figures/appendix/yang_and_mandt.tex}
        \vspace{0.01cm}
        \caption{Additional example comparisons with Yang \& Mandt \cite{yang2022diff}.}
    \end{figure}

    \newpage

    \section{Additional reconstructions}

    \begin{figure}[H]
        \hspace{-1cm}
        \includegraphics[width=9cm]{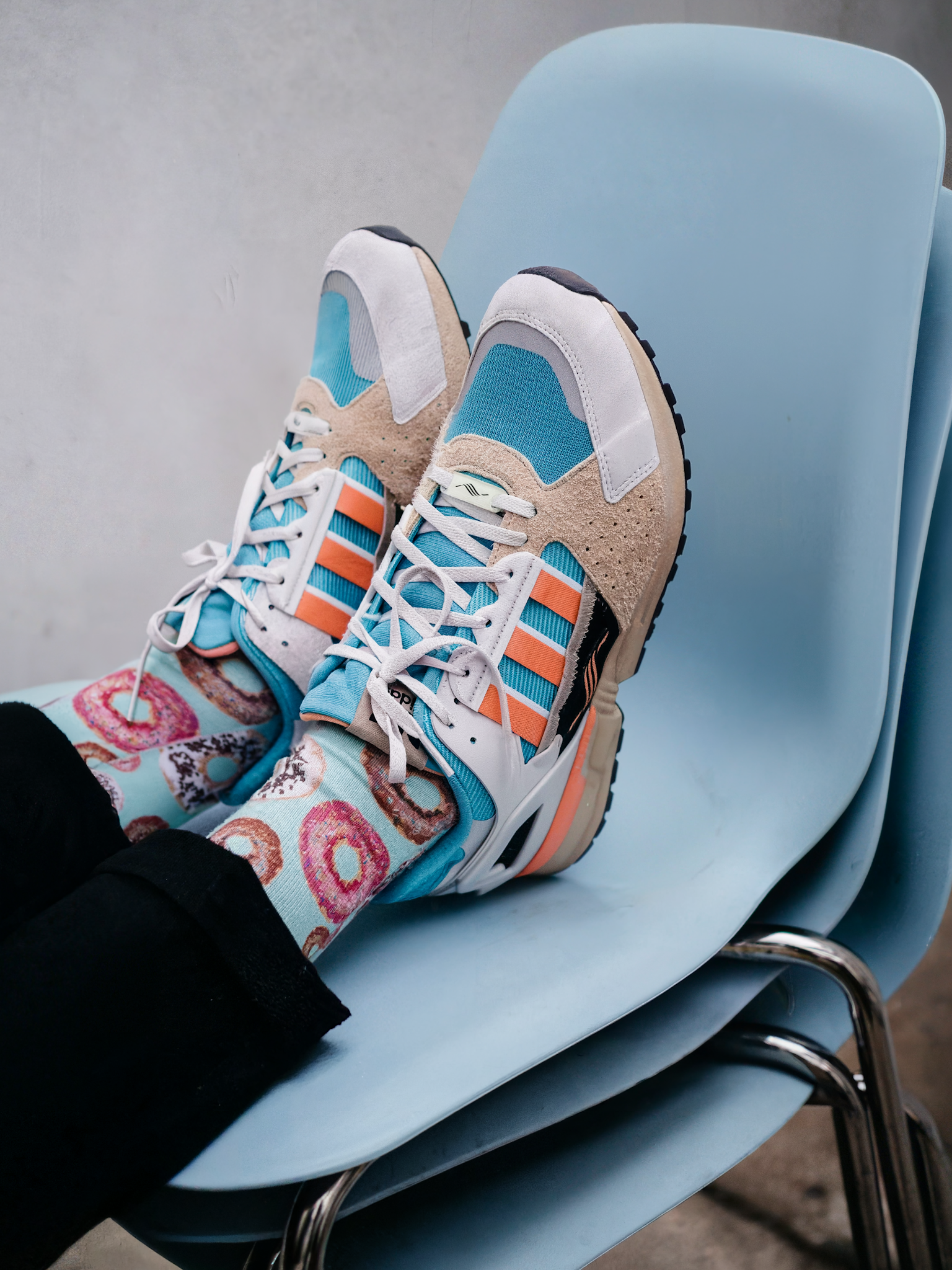}
        \includegraphics[width=9cm]{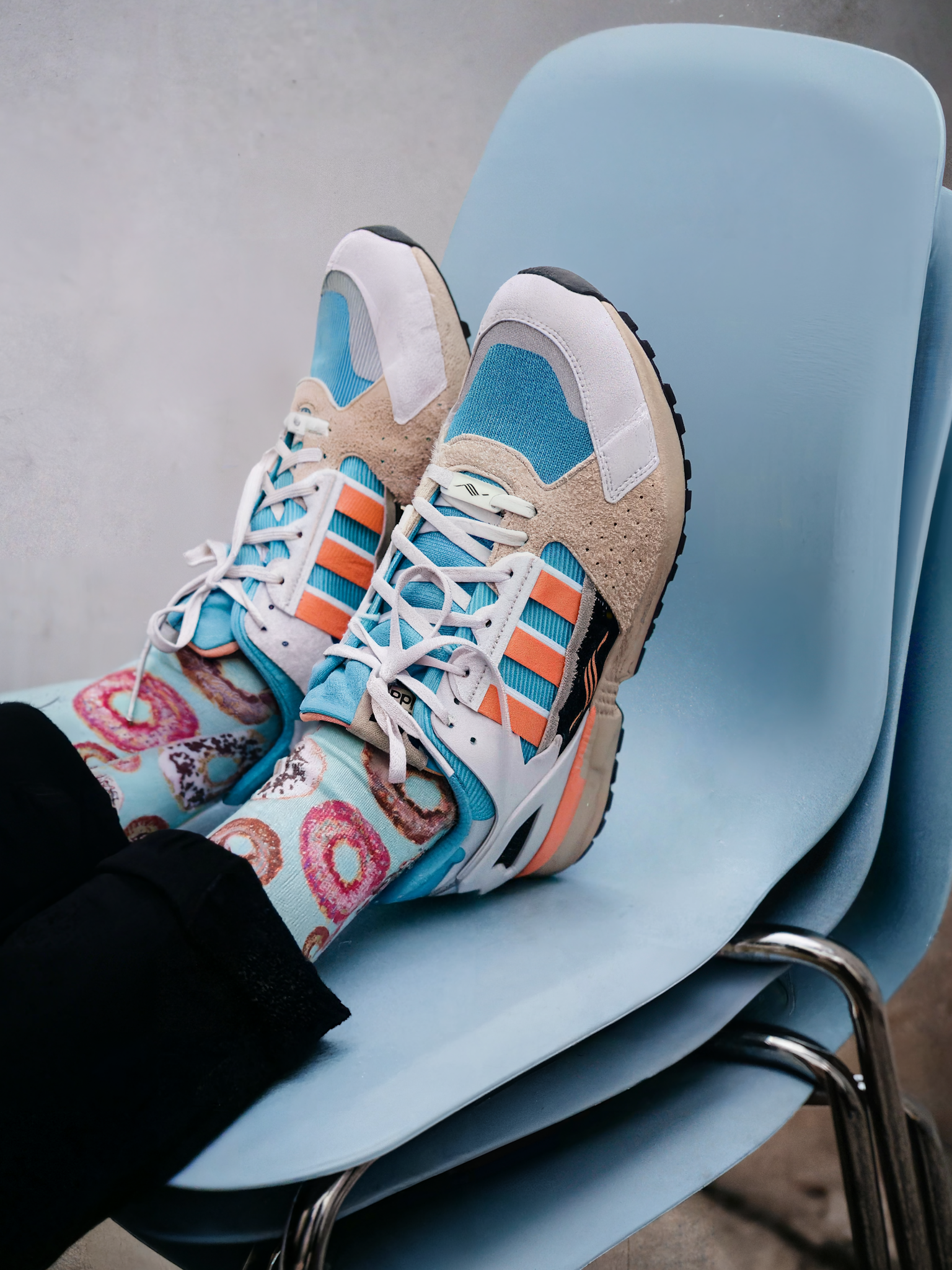}
        \caption{Reconstructions of an image from the CLIC2020 dataset compressed with HFD at 0.0538 bpp (left) and 0.0307 bpp (right), respectively.}
    \end{figure}

    \section{Partial generation}

    \begin{figure}[H]
        \centering
        \includegraphics[width=5cm]{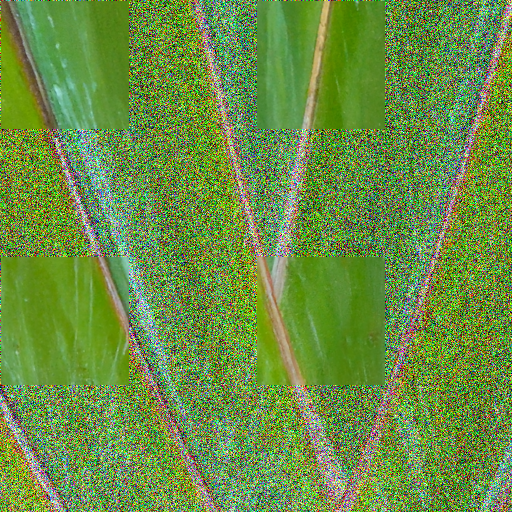}
        \caption{Patches are generated in stages. In the image above, noise-free pixels of four patches have been generated conditioned on noisy pixels in the surrounding patches.}
    \end{figure}

    \section{HFD+}

    \begin{figure}[H]
        \centering
        \includegraphics[height=3.8cm]{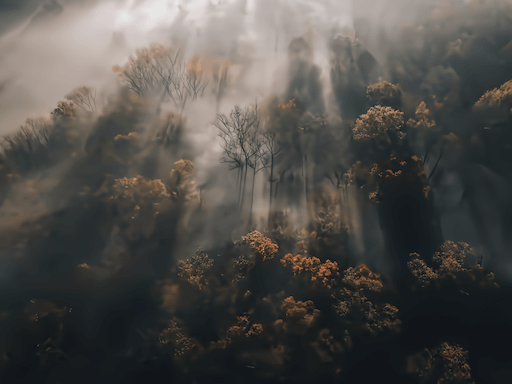}
        \includegraphics[height=3.8cm]{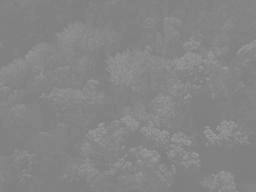}
        \includegraphics[height=3.8cm]{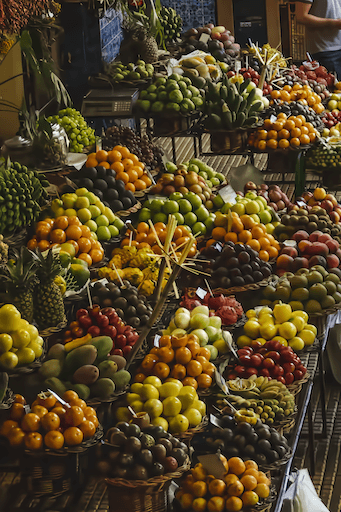}
        \includegraphics[height=3.8cm]{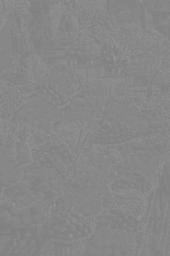}
        \caption{Visualization of inputs to HFD+. The absolute value of residuals, $|\hat{\vx}^{\mathrm{MSE}} - \vx|$, is downsampled by a factor of 8 and then encoded as a JPEG at a very low bit- rate. This residual energy image is fed into the generative model alongside $\hat{\vx}^{\mathrm{MSE}}$.}
    \end{figure}

    \begin{figure}[H]
        \centering
        \input{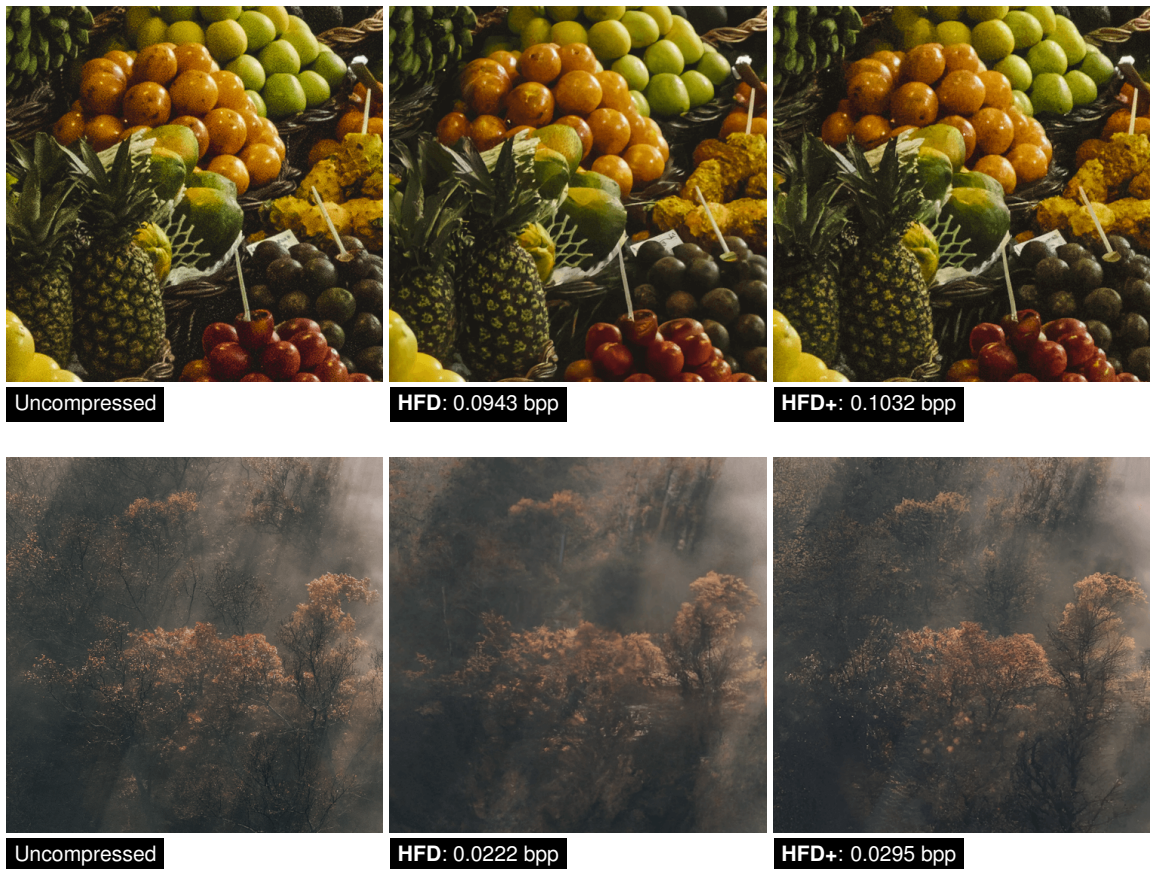}
        \caption{HFD can have a denoising effect (top row). While the result looks pleasing, this effect may not always be desired. Additionally conditioning on the residual energy allows HFD+ to produce a grainier reconstruction which is closer to the uncompressed image. Similarly, HFD is sometimes unable to distinguish between images which were out of focus or whose high frequencies have merely been lost in the reconstruction $\hat{\vx}^{\mathrm{MSE}}$. Conditioning on the residual energy allows HFD+ to hallucinate an appropriate amount of high frequencies (bottom row).}
    \end{figure}

   \section{Architecture}

    \begin{table}[H]
        \centering
        \caption{HFD U-Net architecture\vspace{2ex}}
        \label{tab:architecture}
        \begin{tabular}{l l l l l l l}
        \toprule
            Level & $256\times$ & $128\times$ & $64\times$ & $32\times$ & $16\times$ &\\ \midrule
            Channels & 128 & 128 & 256 & 256 & 1024 \\
            Blocks & 2 & 2 & 2 & 2 & 16 \\
            Attention & - & - & - & - & \checkmark \\ \bottomrule
        \end{tabular}
    \end{table}

    {
    \small
    \putbib
    }
\end{bibunit}

\end{document}